\documentclass[prd,twocolumn]{revtex4}
\usepackage{amssymb}
\usepackage{color}
\usepackage{graphicx}
\usepackage{comment}

\def\gr{$\gamma$-ray}

\begin{document}

\title{IceCube Sensitivity for Neutrino Flux from Fermi Blazars in Quiescent States}
\author{A. Neronov$^{1,2}$ and M. Ribordy$^3$} 
\affiliation{$^1$ ISDC Data Centre for Astrophysics, 
Chemin d'Ecogia 16, CH-1290 Versoix, Switzerland \\
$^2$ Geneva Observatory, 51 ch. des Maillettes, 
CH-1290 Sauverny, Switzerland \\
$^3$ High Energy Physics Laboratory, EPFL, CH-1015 Lausanne, Switzerland}

\begin{abstract}
We investigate the IceCube detection potential of very high energy neutrinos from blazars, for different classes of "hadronic" models, taking into account the limits imposed on the neutrino flux by the recent {\it Fermi} telescope observations. 

Assuming the observed \gr\ emission is produced by the decay of neutral pions from proton-proton interactions, 
the measurement of the time-averaged spectral characteristics of blazars in the GeV energy band 
imposes upper limits on the time-averaged neutrino flux. Comparing these upper limits to the 
$5\sigma$ discovery threshold of IceCube for different neutrino spectra and different source locations in the sky, 
we find that several BL Lacs with hard spectra in the GeV band are within the detection potential of IceCube. 
If the \gr\ emission is dominated by the neutral pion decay flux, none of the flat-spectrum radio quasars are detectable with IceCube. 
If the primary high energy proton spectrum is very hard and/or neutrinos are produced in proton-photon, rather than proton-proton reactions, the upper limit on the neutrino flux imposed by the measured \gr\ spectra is relaxed and gamma-ray observations impose only lower bounds on the neutrino flux. We investigate whether these lower bounds guarantee the detection of blazars with very hard neutrino spectra (spectral index $\Gamma_\nu\sim 1$), expected in the latter type model.

We show that all the "hadronic" models of activity of blazars are falsifiable with IceCube. Furthermore, we show that models with \gr\ emission produced by the decay of neutral pions from proton-proton interactions can be readily distinguished from the models based on proton-gamma interactions and/or models predicting very hard high energy proton spectra  via a study of the distribution of spectral indices of \gr\ spectra of sources detected with IceCube.

\end{abstract}

\maketitle
 
\section{Introduction}

The broadband, radio-to-\gr\ spectra of blazars are most commonly interpreted within the framework of synchrotron-self-Compton (SSC) or synchrotron-external-Compton (SEC) models \cite{maraschi92,dermer93,sikora94,ghisellini96,fossati98}. Within this framework, the observed \gr\ emission is
generated via inverse Compton (IC) scattering of low energy background photons. The background photons originate either from an external source (in SEC models) or are produced by the same high energy electrons via the synchrotron mechanism (SSC models).

An alternative interpretation of the \gr\ emission is considered in models involving interactions of high energy protons \cite{sikora87,mannheim92,aharonian00a}.  Several types of such "hadronic" interaction models are considered in the literature. Different models adopt different assumptions about the dominant interaction channel of high energy protons and about the high energy proton spectrum. 

If high energy protons interacting with low energy protons is the dominant channel (e.g. from the accretion flow onto a supermassive black hole), a significant contribution to the observed \gr\ flux could be secondary emission from the decay of neutral pions, $\pi^0\rightarrow\gamma\gamma$. 
Indeed, if one assumes a "conventional" $E^{-2}$ power-law type spectrum of high energy protons (see e.g. \citep{halzen05}) generated for instance by proton shock acceleration in the blazar jet, one finds that the spectrum of \gr\  produced by the $\pi^0$ decay should be distributed over a broad range of energies, starting from a fraction of the $pp$ reaction energy threshold, $E_{pp,\rm thr}=2m_\pi(1+m_\pi/m_p)\simeq  280$~MeV, where $m_p,m_\pi$ are proton and pion masses, up to the high energy cutoff of the proton spectrum.
In contrast, if the high energy protons interact mostly with the background photons, the reaction threshold is kept high,
$E_{p\gamma,\rm thr}=(m_pm_\pi+m_\pi^2)/\epsilon\simeq 6.8\times 10^{16}\left[\epsilon/1\mbox{ eV}\right]^{-1}$~eV, where $\epsilon$ is the energy of the background photon. 
Consequently, the observed \gr\ emission from blazars, detected at energies $E_\gamma\ll E_{p,\rm thr}$, cannot be produced by neutral pion decay. 
Instead,  the observed \gr\ emission in this type of models is usually supposed to be produced by secondary electron-positron pairs generated in proton interactions and/or in electromagnetic cascade initiated by the proton interaction.

A similar "proton-initiated cascade" origin of the observed \gr\ emission is assumed in models which consider the possibility of very hard spectra of high energy protons, or proton spectra with low energy cutoffs (see e.g. \citep{lovelace76,lesch92,kardashev95,neronov02,rieger08,neronov09,neronov09a}). Such hard proton spectra can be produced if protons are accelerated by large-scale electric fields induced close to the black hole horizon, rather than at shocks in the blazar jets. Arguments in favor of particle acceleration directly in the central engine of active galactic nuclei (AGN) are given by the recent observations of very fast variability of high energy \gr\ emission from blazars and radio galaxies (see e.g.\citep{neronov07a,rieger08,neronov09}).

A straightforward distinction between hadronic and leptonic models and between different types of hadronic models of emission from
blazars would follow from the detection of high energy neutrinos, expected to accompany \gr\ emission in hadronic models only. 
However, experimental limitations have prevented positive outcomes until recently.
First, the sensitivity of existing high energy neutrino detectors was not sufficient to detect  neutrino flux even from the brightest blazars. Search results for neutrino signal from astronomical point sources have until now only provided upper limits \cite{Desai:2007ra,Abe:2006at,amanda_ps} usually above model predictions. Next, the quality of the available \gr\ data was not sufficient to make firm predictions of the time-averaged (over flaring and quiescent states, taking into account flaring "duty cycles") neutrino fluxes from blazars expected in different hadronic emission models. 

With the newly operating IceCube \cite{icecube} and {\it Fermi} \cite{Atwood:2009ez} telescopes, a new sensitivity window on high energy neutrinos is currently opening (see e.g. ~\cite{Gaisser:1994yf} where the IceCube science potential has been generically described or ~\cite{Halzen:2008zj,Halzen:2007ah} regarding galactic sources) and the detection of quiescent (time-averaged) levels of \gr\ emission from a large number of blazars has become possible. 
In this paper we discuss the implications of the time-averaged \gr\ spectral measurements of a large number of blazars for the detection of neutrinos from this class of sources by IceCube.


Under the assumption of an observed \gr\ emission in the GeV energy band dominated by neutral pion decay, the time averaged neutrino flux from blazars can be predicted. 
Assuming a simple power-law type spectrum of the primary high energy protons, the measurement of the $\pi^0$ decay flux and spectral index in the GeV band fully constrains the expected neutrino flux at higher energies. 
Using the first published {\it Fermi} catalogue of blazars and AGN (normalization and spectral characteristics) \cite{abdo09}, the corresponding neutrino fluxes for each blazar are calculated.
Once these neutrino fluxes convolved to the instrumental response of IceCube for sources at different declinations, the expected number of neutrinos for each blazar is derived. 
This analysis leads us to conclude a detectable steady-state neutrino emission from a few (out of $\sim 10^2$) {\it Fermi} blazars with the hardest spectra in the GeV band.

If the observed \gr\ emission is dominated by the synchrotron and/or inverse Compton and/or Bremsstrahlung emission from the secondary $e^+e^-$ pairs produced in proton-initiated particle cascades, the constraints on the neutrino flux from the source are significantly weaker, 
since the measurement of the GeV band \gr\  spectrum does not provide a direct constraint on the neutrino spectrum. We demonstrate this possibility by considering a model in which a nearly monoenergetic proton beam (composed of e.g. protons accelerated in a vacuum gap in a black hole magnetosphere) produces \gr s and neutrinos in proton-proton interactions (e.g. during the propagation through the accretion flow). 

By combining IceCube and {\it Fermi} data, we show that it should be possible to (a) confirm or disprove the hadronic models of activity of blazars and (b) firmly distinguish between models which predict very hard neutrino spectra (which could be either due to the low energy cutoffs in the proton spectra or due to the threshold effects of the proton-photon interactions) and models which assume relatively soft ($E^{-2}$-like) neutrino spectra). This would proceed via a study of the distribution of neutrino-detected blazars in a \gr\ "Flux vs. photon index" diagram in the GeV band: 
In the scenario with neutral pion decay \gr\ emission from proton-proton interactions all the blazars detected with IceCube should occupy the "high flux - hard spectrum" corner of the diagram, 
while they should be randomly distributed in the models based on $p\gamma$ interactions and/or proton spectra with low energy cutoffs.

The plan of the paper is as follows: In Section \ref{sec:neutralpion} we consider a specific hadronic model of blazar activity which assumes \gr\ emission produced a power-law distribution of high energy protons (with a relatively soft spectra) in interaction with low energy protons from the accretion flow.  
We show that in this class of models the spectral characteristics of the GeV-band \gr\ emission are directly related to the spectral characteristics of neutrino emission. 
In Section \ref{sec:fermi} we derive 
the time-averaged neutrino flux from blazars detected in the GeV energy band by the {\it Fermi} telescope
and demonstrate that only blazars with the hardest GeV-band spectra can be expected to be detectable as neutrino sources.

Next, in Section \ref{sec:hard} we present a specific hadronic model which 
predicts a very hard neutrino spectrum. 
We demonstrate that in this class of models, measurements of the GeV band \gr\ spectra by {\it Fermi} impose only lower bounds on the neutrino flux, but do not constrain the spectral characteristics of the neutrino signal. 
We discuss the numerous uncertainties affecting the neutrino flux estimates. 
In Section \ref{sec:sensitivity} we investigate whether blazars with very hard neutrino spectra can be detected with IceCube. One immediate problem for the detection of sources with very hard spectra is related to the Earth opacity to neutrinos with energies above $\sim$PeV, which dominate the source flux. The statistics of the neutrino signal at increasing source declinations is therefore reduced. We then derive the IceCube spectral discovery curves for arbitrary source declinations. Finally, in Section \ref{sec:hadronic} we investigate whether the hadronic model(s) of blazar activity are falsifiable with IceCube and whether different types of hadronic models can be distinguished using a combination of IceCube and {\it Fermi} data.

\section{GeV \gr\ emission from neutral pion decay}
\label{sec:neutralpion}

If the high energy protons are produced via shock acceleration, their spectrum is expected to be roughly a cutoff powerlaw with a spectral index $\Gamma_p\sim 2$,  $dN_p/dE\sim E^{-\Gamma_p}\exp(-E/E_{p,\rm cut})$. 
If collision of high energy protons with low energy protons is the main interaction channel (e.g. from the accretion flow onto supermassive black hole, or with protons populating the blazar jet), both the $\pi^0$ decay \gr\  and the neutrino spectra are cutoff power-laws with approximately  the same indices, $\Gamma_\nu \simeq \Gamma_\gamma \simeq \Gamma_p$, in a wide energy interval from $\sim 0.1$~GeV up to a cutoff at energies somewhat smaller than that of the primary proton spectrum $E_{\gamma,\nu,\rm cut}\simeq KE_{p,\ \rm cut}$ with $K\sim 0.1$ \cite{aharonian_book}. 

\begin{figure}
\includegraphics[width=\linewidth]{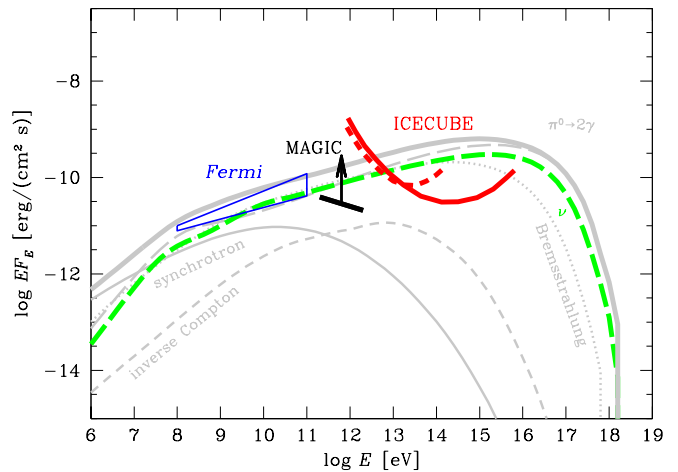}
\caption{Spectra of \gr\  and neutrino emission of blazar PG 1553+11 expected in a model in which the GeV emission is dominated by the \gr s from neutral pion decay. The thin solid, short dashed and dotted lines show the contributions from synchrotron, inverse Compton and Bremsstrahlung emission from the secondary $e^+e^-$ pairs. The two curves marked IceCube show the discovery potential of IceCube to the power-law like neutrino spectra with different spectral indices, for different source declinations: $15^\circ$ (thick solid red curve) and $75^\circ$ (thick dashed red curve), derived in Section \ref{sec:sensitivity}.}
\label{fig:e-2}
\end{figure}

Naively, the similarity between the \gr\ and neutrino spectra should enable to predict the blazar neutrino emission characteristics if the \gr\ emission spectrum is known. 
The IceCube telescope being most sensitive to neutrinos with energies above $\sim 1$~TeV, the knowledge of the \gr\ source spectrum in this energy band is most important to estimate the detectability of the neutrino signal.
However, in the case of extragalactic sources, the \gr\  flux at energies above $\sim 10$~GeV -- 1~TeV is attenuated by pair production on the background radiation fields (collectively known as Extragalactic Background Light, EBL). This degrades the information carried by the shape and overall flux of the $\pi^0$ decay spectrum of the source. This, in turn, prevents a direct assessment of source detectability with a neutrino telescope. Even for relatively low redshift sources, such as TeV \gr\ -loud blazars, measurements of source spectra above 100 GeV do not provide information about the source neutrino spectra. This is illustrated in Fig. \ref{fig:e-2}, in which we show a model fit for the GeV-TeV band  emission spectrum of the blazar PG 1553+11, recently measured by the {\it Fermi}, HESS and MAGIC telescopes. The TeV band observations result only in a lower bound on (rather than measurement of) the source \gr\ flux (see e.g. \citep{persic08}). 

Contrary to TeV \gr s, \gr s in the GeV energy band do not suffer from the absorption on the EBL. Measurements of the spectral characteristics in this energy band provide, therefore, information about the intrinsic spectrum of the blazar. Since the $\pi^0$ decay spectrum is expected to extend down to sub-GeV energies in models based on proton-proton interaction, a measurement of the spectral characteristics in the GeV energy band could be used to constrain the $\pi^0$ decay spectrum. Assuming a power-law primary proton spectrum, the intrinsic \gr\ spectrum of the source in the TeV energy range can be reconstructed and the neutrino flux from the measurements of the flux and spectral index of the \gr\ emission in the GeV band can be estimated.

This possibility is illustrated in Fig. \ref{fig:e-2}, in which we show the results of a numerical calculation of the \gr\ and neutrino spectra produced in proton-proton interaction of high energy protons with a power-law spectrum with index $\Gamma_p=1.8$ and a high energy cutoff at $E_{p,\ \rm cut}=10^{18}$~eV, with low energy protons from the (radiatively inefficient) accretion flow onto a $10^7M_\odot$ black hole. Our model is identical to the model of high energy emission from the cascade developing at the base of the jet of a TeV blazar, PKS 2155-304, developed in the Ref. \citep{neronov09}. Protons are initially accelerated close to the black hole horizon (at the distance $D\sim 10^{12}$~cm) and escape to large distances (our calculation stops at $D=10^{16}$cm). Interaction with protons of the accretion flow (see \citep{neronov09} for the detailed description of the geometry of the radiatively inefficient accretion flow) initiate a particle cascade which accompanies the escape of the high energy protons. The cascade leads to the generation of neutrino and \gr\  emissions, calculated using the analytical approximations for the production spectra of secondary particles in $pp$ interaction derived by \citet{kelner06}. 

From Fig. \ref{fig:e-2} one can see that the GeV band emission is dominated by (or, at least, has a significant contribution from) the $\pi^0$ decay emission component, which has a spectral energy distribution very similar to the one from neutrinos. 
This enables an estimate of the characteristics of neutrino spectrum from the GeV-band \gr\ spectrum.

One should note that, apart from the $\pi^0$ decay emission, the \gr\ spectrum in the GeV band could have contributions from  synchrotron, inverse Compton and Bremsstrahlung emission produced by the secondary $e^+e^-$ pairs. Such $e^+e^-$ pairs are produced by charged pion decay and/or in result of the development of electromagnetic cascade inside the blazar jet. In the pure hadronic model of blazar activity, where all the high energy electrons and positrons are supposed to be produced in this way, the power injected in electrons/positrons from the $\pi^\pm$ decay is at most comparable to the power injected into $\pi^0$ decay \gr s. If the primary proton spectrum is relatively soft, $\Gamma_p \sim 2$, the power transferred to the $e^+e^-$ pairs by the electromagnetic cascade is also at most comparable to the initial $\pi^0$ decay \gr\ power. 
This means that in models based on proton-proton interaction with soft high energy proton spectra, 
a significant contribution in the GeV band \gr\ emission is given by the $\pi^0$ decay component, in spite of the possible presence of synchrotron, inverse Compton and/or Bremstrahlung emission components. This ensures the possibility to estimate the neutrino signal from the detected GeV \gr\ signal.

\section{Detectability of quiescent neutrino emission from blazars in the light of Fermi}  
\label{sec:fermi}

Measurements of the spectra of quiescent emission from blazars at energies below 10 GeV became possible recently with the start of operation of the {\it Fermi} telescope. Measurements of the spectral parameters of the \gr\  emission from blazar, reported in \citep{abdo09} enables, for the first time, to make predictions of the time-averaged neutrino flux within the hadronic model of activity, under the assumption that the GeV band flux is dominated by the $\pi^0$ decay contribution.

\begin{figure}
\includegraphics[width=\linewidth]{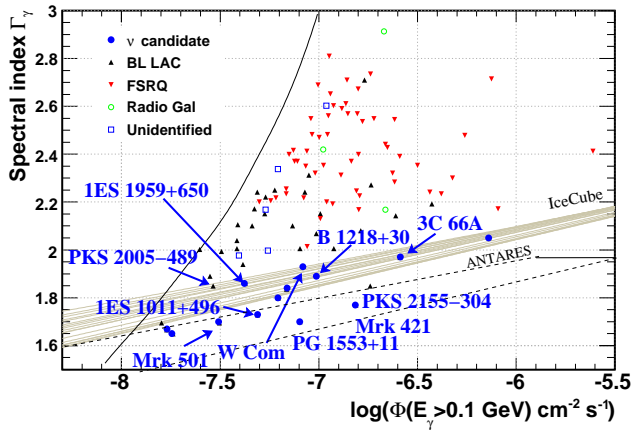}
\caption{Distribution of {\it Fermi} AGN in "flux vs. spectral index" parameter space (from \citet{abdo09}).  Grey curves show the $5\sigma$ "discovery" threshold for detection of neutrino signal from {\it Fermi} blazars by IceCube in the model in which \gr\ emission is dominated by the signal from neutral pion decay. Different curves correspond to different assumed source declinations, from $\delta=5^\circ$ (upper curve) to $\delta=85^\circ$ (lower curve) with a $\Delta\delta=5^\circ$ step. Below the 3-yr IceCube discovery curves are indicated the 1-yr ANTARES 90\% C.L upper limit curves, extracted and extrapolated from \cite{Kalekin:2009zz} (see text).  
Blue filled circles correspond to {\it Fermi} blazars considered as "neutrino source candidates" in the model with neutral pion decay \gr\ emission.}
\label{fig:diagram}
\end{figure}

The resulting estimates of the number of neutrinos from different blazars, given three years of IceCube exposure (in the 80 string baseline configuration) \cite{icecube} are shown in Table \ref{tab:blazar_list} (only blazars with an expected number of neutrinos close or above the $5\sigma$ ''discovery'' threshold are listed).


The $5\sigma$ ''discovery'' threshold is calculated using the declination-dependent atmospheric neutrino background parametrization from \cite{volkova} (which includes the ``prompt'' $\nu_\mu$ flux). The ``prompt'' component is as yet uncertain and may significantly contribute to the overall atmospheric neutrino spectrum at energies above $E\sim 10^{5-6}$~GeV. 

The diffuse neutrino emission from the interaction of galactic cosmic rays with the interstellar medium, estimated for instance in Ref. \cite{galcr}, has been neglected: While expected to be relatively important for sources at low galactic latitudes $b\sim 0^\circ$, but not exceeding the ``prompt'' atmospheric neutrino component, it is largely sub-dominant at large galactic latitudes.

The diffuse neutrino emission from unresolved blazars was neglected as well. Within the proton-proton interaction models considered above and based on early measurements of the EGRET extragalactic diffuse \gr\ background by \cite{strekumar}, the diffuse extragalactic neutrino flux was estimated to eventually dominate the atmospheric neutrino flux above $E\sim 10^{5}$~GeV \cite{dar}.
However, revised measurements of the extragalactic diffuse \gr\ background with EGRET indicate a lower normalization and a softer background spectrum \cite{strong04}, thus accordingly reducing this neutrino background component from a rescaling of the estimates in Ref. \cite{dar} to a level well below the atmospheric neutrino flux at $\sim 10^5$~GeV.

The uncertainty of the diffuse neutrino background level at energies above $E\sim 10^{5}$~GeV may slightly modify the results of Table \ref{tab:blazar_list}, especially the numbers of neutrino-induced muon events with detector threshold energy $E_{\mu}^{min,2}=10^{4.5}$~GeV. On the other hand, the numbers of neutrino-induced muon events with detector threshold energy $E_{\mu}^{min,1}=10^{3.5}$~GeV remain almost unaffected, the dominant diffuse background source being the conventional atmospheric neutrino component.


To estimate the parameters of the neurino spectrum (normalization $\Phi_\nu$, and spectral index $\Gamma_\nu$) from the observed parameters of the \gr\ spectrum ($\Phi_\gamma, \Gamma_\gamma$), we produce a "lookup table" $(\Gamma_\nu , \Phi_\nu )$ vs. $(\Gamma_\gamma , \Phi_\gamma )$ using the neutrino and \gr\ production spectra for different powerlaw distributions of the primary protons. 
We use the approximations of \citet{kelner06} for the calculation of the production spectra and assume a fully oscillated neutrino flux ({\it i.e.} the normalization of the muon neutrino spectrum is $1/3$ of the normalization of the primary total neutrino spectrum).  
The number of detected neutrinos from the source (and the contamination by the atmospheric neutrino background) depends on the low muon energy threshold chosen for the analysis. In Table~\ref{tab:blazar_list}, we show the number of neutrinos from the source for two different choices of the reconstructed muon energy thresholds,
$E_\mu^{min,1}=10^{3.5}\,\mathrm{GeV} \simeq 3\,\mathrm{TeV}$ and $E_\mu^{min,2}=10^{4.5}\,\mathrm{GeV} \simeq 30\,\mathrm{TeV}$.
In analyzing the data, the best choice for the reconstructed muon energy threshold depends on the neutrino spectral index, $\Gamma_\nu$. 
This issue is discussed in greater details in Section \ref{sec:sensitivity}.

Out of 104 bright blazars detected by {\it Fermi} during the first three months of operation \citep{abdo09}, only a few are expected to produce sufficiently high neutrino flux for a detection within one year of IceCube operation. All the neutrino candidate blazars are BL Lac type objects. None of the bright and powerful Flat Spectrum Radio Quasars (FSRQ), such as 3C 279, 3C 273 or 3C 454.3, are expected to produce a neutrino flux above the IceCube detection threshold.

\begin{table*}
\begin{tabular}{lllllll}
\hline
Name & {\it Fermi} Name &$z$ & $\Gamma_\gamma$ & $F$ & $(n_s, n_b, n_{5\sigma})_{E_\mu^{min,1}}$ & $(n_s, n_b, n_{5\sigma})_{E_\mu^{min,2}}$ \\
\hline
\multicolumn{7}{c}{Blazars above $5\sigma$ detection threshold}\\
\hline
B3 0133+388 &0FGL J0136.6+3903 & (?) & 1.65 & 1.8 & (14, 0.36, 7) & (4.6, 0.0079, 3) \\
{\bf 1ES 1011+496} &0FGL J1015.2+4927 & 0.212 & 1.73 & 4.9 & (12, 0.31, 7) & (3.1, 0.0067, 3)\\
{\bf Mrk 421} &0FGL J1104.5+3811 & 0.0300 & 1.77 & 15.3 & (27, 0.36, 7) & (7.7, 0.0080, 3)\\
PKS 1424+240 &0FGL J1427.1+2347 & (?) & 1.80 & 6.2 & (11, 0.53, 8) & (3.6, 0.013, 3)\\
{\bf PG 1553+11 }&0FGL J1555.8+1110 & 0.36(?) & 1.70 & 8.0 & (71, 0.95, 10) & (31, 0.024, 4)\\
{\bf Mrk 501} &0FGL J1653.9+3946 & 0.0337 & 1.70 & 3.1 & (13, 0.35, 7) & (3.9, 0.0077, 3)\\
PKS 1717+177 &0FGL J1719.3+1746 & (?) & 1.84 & 6.9 & (7.9, 0.67, 8) & (2.8, 0.017, 4)\\
\hline
\multicolumn{7}{c}{Blazars close to $5\sigma$ detection threshold}\\
\hline
{\bf 3C 66A} &0FGL J0222.6+4302 & 0.444 & 1.97 & 25.9 & (3.3, 0.34, 7) & (0.74, 0.0076, 3) \\
AO 235+164 &0FGL J0238.6+1636 & 0.94 & 2.05 & 72.6  & (5.2, 0.70, 9) & (1.5, 0.018, 4)\\
1ES 0502+675 &0FGL J0507.9+6739 & 0.416 & 1.67 & 1.7 & (4.8, 0.24, 6) & (0.91, 0.0038, 3)\\
{\bf B 1218+30 }&0FGL J1218.0+3006 & 0.182 & 1.89 & 9.7 & (4.3, 0.43, 7) & (1.2, 0.010, 3)\\
{\bf W Com} &0FGL J1221.7+2814 & 0.102 & 1.93 & 8.3 & (2.3, 0.46, 7) & (0.63, 0.011, 3)\\
{\bf 1ES 1959+650} &0FGL J2000.2+6505 & 0.047 & 1.86 & 4.2 & (1.3, 0.24, 6) & (0.20, 0.0041, 3)\\
\hline
\multicolumn{7}{c}{Neutrino candidate blazars in Southern hemisphere}\\
\hline
\textcolor{blue}{KUV 0311-1938} &0FGL J0033.6-1921 & 0.610 & 1.70 & 1.6 & &\\
\textcolor{blue}{\bf PKS 2005-489} &0FGL J2009.4-4850 & 0.071 & 1.85 & 2.9 & &\\
\textcolor{blue}{\bf PKS 2155-304} &0FGL J2158.8-3014 & 0.116 & 1.85 & 18.1 & &\\
\hline
\end{tabular}
\caption{List of blazars with potentially detectable quiescent neutrino emission, in the model with \gr\ flux in the {\it Fermi} band dominated by neutral pion decay. Bold names: blazars detected in the TeV energy band. Blue (Grey) names: blazars in the Southern hemisphere, not accessible to IceCube. The two last columns list the expected number of detected signal and atmospheric neutrinos, $n_s$ and $n_b$, together with the estimated $5\sigma$ discovery threshold, $n_{5\sigma}$ for two choices of the reconstructed low muon energy threshold.
$E_\mu^{min,1}$ and $E_\mu^{min,2}$ (see text).
}
\label{tab:blazar_list}
\end{table*}

This is also clear from Fig. \ref{fig:diagram} where we show the thresholds of detectability  of neutrino emission from {\it Fermi} blazars, calculated by normalizing the neutrino flux from the {\it Fermi}  \gr\ spectra.  In principle, the expected number of high energy neutrinos from the source depends not only on the intrinsic source spectrum, but also on the position of the source in the sky (given the specific location of IceCube at the South Pole, on the source declination). Not only the number of source neutrinos, but also the atmospheric neutrino flux, which constitutes the background for the source signal, depend on the declination. Thus, it follows a dependence of the $5\sigma$ source detection threshold on the declination. Fig. \ref{fig:diagram} shows that the best candidates for a detection with IceCube (within proton-proton models with soft high energy proton spectra) are the low declination BL Lacs with hard GeV band \gr\  spectra. All the bright FSRQ detected by {\it Fermi} are expected to produce neutrino counts below the detection threshold of IceCube, independent of their declination. 

To understand why the bright FSRQ are not detectable by IceCube, one has to take into account that the predictions for the neutrino flux depend not only on the normalization of the \gr\  spectrum at GeV energies, but on its hardness as well. IceCube is most sensitive to neutrinos with energies above $\sim 1$~TeV. The harder the spectrum of the source in the GeV energy band, the higher the expected flux in the TeV-PeV band. It turns out that in the GeV band, FSRQ have spectra on average softer than BL Lacs and thus explains why the less luminous (in the GeV band) BL Lacs are expected to produce higher neutrino fluxes in the TeV-PeV band than FSRQs. 
These estimates rely however on an initially foreseen 80 string IceCube configuration and enhanced sensitivity to lower energy in the new 86 string baseline configuration has not been considered yet \cite{Cowen:2008zz}. 

It is worth mentioning that most of the best candidate sources listed in Table \ref{tab:blazar_list} are TeV \gr\  loud BL Lacs. 10 out of 16 sources have already been reported to be TeV emitters. The remaining 6 have either large or unknown redshift which explains the difficulty of their detection above 100 GeV. It is unlikely that the TeV blazar neutrino emission be produced in $p\gamma$ interaction, because a high efficiency of the $p\gamma$ interaction (which has a cross section $\sigma_{p\gamma}\sim 10^{-28} $~cm$^2$) would automatically imply a strong absorption of the TeV \gr\ s via pair production on the soft photon background in the source (the cross section for this process is $\sigma_{\gamma\gamma}\simeq 10^{-25}$~cm$^2$, {\it i.e.} three orders of magnitude larger than the $p\gamma$ cross section) \citep{aharonian00,neronov02}. The only possibility for neutrino production in TeV blazars is via proton-proton interactions.

The detection potential of ANTARES for the strongest neutrino emitters can also be roughly estimated using the one year $E^{-2}$ 90\% C.L upper limit curves found in Ref. \cite{Kalekin:2009zz} ($E^2 {{\rm{d}}\Phi_{\nu_\mu}}/{{\rm{d}}E_\nu} \approx 4\,-\,10 \times 10^{-8} \,\rm{GeV}\,\rm{cm}^{-2}\,\rm{s}^{-1}$ over the declination range $-90^\circ < \delta < 30^\circ$), assuming a behavior similar to IceCube for hardening spectra  ({\it i.e.} by extrapolating parallel to IceCube 5$\sigma$ discovery curves). The expected signal escapes ANTARES discovery potential for all sources but for PG 1553+11, which could be in the reach of ANTARES after one year. Note however that this assumption may not hold, see Appendix \ref{sec:append}, given the distinct detector response of ANTARES and IceCube.

\section{Hadronic models with very hard neutrino spectra}
\label{sec:hard}

The above IceCube neutrino count estimates from blazars show that within the hadronic models predicting relatively soft neutrino spectra ($E^{-2}$ like),  $\le 10$  \gr\  loud blazars with hardest GeV band \gr\ spectra can be detected within a several year-long neutrino data set. 

Although experimental results in the literature usually consider benchmark power-law type neutrino spectra $dN_\nu/dE\propto E^{-\Gamma_\nu}$ with $\Gamma_\nu\sim 2$
most existing hadronic models of blazar activity predict neutrino spectra which significantly differ from a $E^{-2}$ powerlaw. 
%
This is often related to the existence of a low energy cutoff in the proton spectra.
As it is mentioned in the Introduction, these low energy cutoffs can be either intrinsic cutoffs in the proton spectra (e.g. in the models with proton acceleration in the blazar central engine) or can arise due to the existence of a threshold energy of proton-photon interactions.

Since the low energy cutoff or the photo-pion threshold energy, considered in most of the hadronic models, is much higher than $\sim 1$~PeV, the contribution of the $\pi^0$ decay component to the  \gr\  emission spectrum in the GeV - TeV energy band is small, compared to the synchrotron and/or inverse Compton and/or Bremsstrahlung emission from the secondary  $e^+e^-$ pairs produced in the decay of the charged pions and/or in electromagnetic cascade. Therefore, the estimates of the expected spectral characteristics of neutrino emission based on the spectral characteristics of the GeV band  \gr\  emission, presented above, do not work.

\begin{figure}
\includegraphics[width=\linewidth]{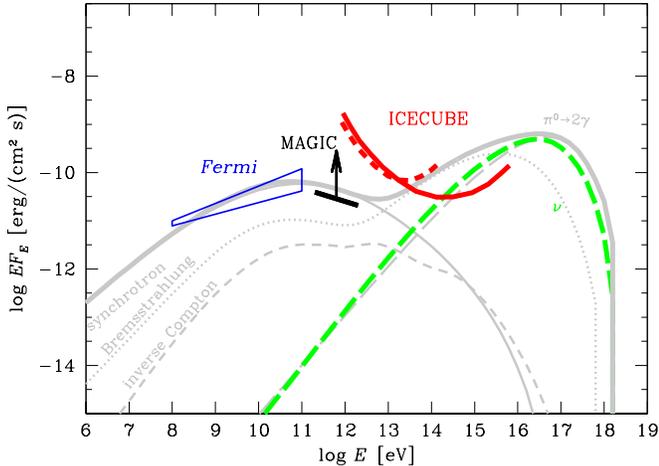}
\caption{Spectra of \gr\  and neutrino emission of blazar PG 1553+11 expected in a model with GeV emission dominated by \gr s from secondary $e^+e^-$ pairs. 
The notations are the same as in Fig. \ref{fig:e-2}.}
\label{fig:e-1}
\end{figure}

To demonstrate this, we show in Fig. \ref{fig:e-1} the result of calculation of the  \gr\  and neutrino spectra in the same model as in Fig. \ref{fig:e-2}, but assuming a very hard high energy proton injection spectrum, with $\Gamma_p \simeq 0$ ({\it i.e.} most of the power of the primary proton beam is carried by protons with energies close to the assumed cutoff energy $E_{p,\ \rm cut}=10^{18}$~eV). One can clearly see that a satisfactory fit of the GeV-TeV  \gr\  spectrum is achieved with the dominant contribution from synchrotron emission from the secondary electrons. No information on the $\pi^0$ decay spectrum, and hence no information about the neutrino spectrum can be extracted from the available  \gr\  data.

The hypothesis of hard neutrino spectra opens a possibility of strong neutrino emission, even in the quiescent state, from a large number of blazars, including both BL Lacs and FSRQs.

It was argued in Ref. \citep{neronov09b} that even if the GeV - TeV band  \gr\  emission is dominated by the flux from electron-positron pairs from $\pi^\pm$ decay, 
constraints on the neutrino luminosity of the source can still be derived from the spectra of the broad band  \gr\  emission. 
Namely, a measurement of the overall bolometric luminosity of a Galactic  \gr\  source imposes a restriction on its neutrino luminosity 
(but not on the spectral index or cutoff energy of the neutrino spectrum).

In the case of extragalactic sources, measurements of the electromagnetic luminosity of the source also imposes a restriction on its neutrino luminosity. However, the relation between the observed  \gr\  power and morphology of an extragalactic source and its neutrino luminosity is more complicated than in the case of a Galactic source.

The problem being that it is possible that even for sources at small redshifts, most of the  \gr\  power initially emitted in the energy band above 0.1 - 1 TeV is absorbed during propagation through the intergalactic medium, due to pair production on the EBL. This means that the primary source power produced in result of proton interaction is not known.  At the same time, to obtain a correct estimate of the neutrino luminosity, one has to be able to estimate the primary source power at energies above 0.1 - 1 TeV. This could, in principle, be done, because the absorbed power is initially converted into  $e^+e^-$ pairs and then back to the  \gr\ s, because the  $e^+e^-$ pairs loose energy via inverse Compton scattering on the CMB photons. Measuring the flux of the inverse Compton emission from the  \gr -induced cascade  $e^+e^-$ pairs one would be able to measure the flux of the initial  \gr\ s and in this way to measure the  \gr\  luminosity of the primary source.

The method of measurement of the flux from the \gr -induced cascade  $e^+e^-$ pairs heavily depends on the strength of the extragalactic magnetic fields.
If the extragalactic magnetic fields are strong, $B\gg 10^{-12}$~G, the trajectories of the  $e^+e^-$ pairs produced in the process of absorption of the  \gr s on the EBL would strongly deviate before they transfer energy through inverse Compton scattering to the CMB photons \citep{aharonian94}. The emission from the  \gr -induced cascade should therefore be detectable as an extended \gr\ halo around the primary point source. The size of the halo is determined by the mean free path of the  \gr\ s through the EBL and thus depends on the primary  \gr\  energies. The luminosity of the halo is equal to the  \gr\  luminosity of the primary source in the TeV - EeV energy band, averaged over $10^3 - 10^6$~yr time scale.

If the extragalactic magnetic fields are somewhat weaker, $10^{-16}\ll B\ll 10^{-12}$~G, the deflection of  $e^+e^-$ pairs over the radiative cooling time would be moderate, but still the signal from the  \gr\  induced cascade in the intergalactic medium could also be detected as an extended emission region around an extragalactic point source \citep{elyiv09,neronov07}, with the source size having a somewhat different dependence on the energy than for halos discussed by \citet{aharonian94}.

Finally, if the magnetic fields in the intergalactic medium are $B\ll 10^{-16}$~G, the extended emission from the  \gr -induced cascade in the intergalactic space is indistinguishable from the point source emission, because the size of the extended source becomes smaller than the point spread function of the current generation  \gr\  telescopes \citep{plaga95}. In this case a measurement of the point source flux provides a good estimate of the total electromagnetic and neutrino power of the source.

One should note, however, that at present
the strength of the extragalactic magnetic fields is largely unknown and
the expected extended  \gr\  emission from the extragalactic sources is not detected.
This means that
the luminosity of the GeV - TeV  \gr\  loud blazars at energies above several TeV is largely unconstrained and, as a consequence, 
the neutrino luminosity of the blazars is not constrained by the  \gr\  data.

The overall luminosity of the source, generated by the interaction of high energy protons is larger or equal to the power contained in electromagnetic cascades, 
which transfer the initial source power to the sub-TeV energy band. 
Since the total electromagnetic luminosity of the source is comparable to its neutrino luminosity,  measurement of the source flux in the sub-TeV energy band provides a lower bound on the neutrino luminosity of the source. Taking this fact into account we show the calculated neutrino flux in Fig. \ref{fig:e-1}, as a lower limit on the expected neutrino flux, obtained under the assumption of an observed \gr\ luminosity of the source comparable to its intrinsic \gr\ luminosity. 

From the discussion of this section it is clear that the search for very hard neutrino sources is extremely important in the context of the study of blazar activity mechanisms. It is not clear {\it a priori}, if neutrino sources with very hard spectra, with most of the energy output in the $> 10$~PeV energy band, are accessible for study with IceCube. The problem being that at energies $>10$~PeV the Earth is not transparent to neutrinos. Thus most of the signal from the source is absorbed in the Earth and does not reach the detector. To the best of our knowledge, no comprehensive study of sensitivity of IceCube to sources with very hard neutrino spectra was published up to now. 
Taking this into account, we investigate in the following section the potential of IceCube for the detection of neutrino sources with very hard spectra extending to the multi-PeV -- EeV energies.

\section{IceCube sensitivity for astronomical sources with hard neutrino spectra}
\label{sec:sensitivity}

The sensitivity of a neutrino telescope to a flux model can be derived from the knowledge of the neutrino effective area of the telescope at the source declination (from the available information in publications, usually a declination-averaged effective area) and the angular \& energy resolution of the instrument together with the assumption of an atmospheric neutrino flux. Nevertheless, this compact approach suffers limitations owing to the fact that the knowledge of the neutrino effective area alone does not permit the derivation of the optimal sensitivity to an arbitrary neutrino flux shape, as detailed information on the detector response is hidden behind the neutrino effective area concept. 

A refined method to assess the potential of a neutrino telescope to an arbitrary flux and arbitrary declination is based on the detector muon effective area and differential muon flux at the detector, which can both be respectively extracted from the averaged neutrino effective area and from the neutrino flux spectrum propagated through the Earth following the methodology introduced by~\citet{neronov09b}. 
From the interplay of neutrino signal contamination by atmospheric neutrinos and neutrino absorption, the optimal reconstructed muon energy threshold for highest sensitivity for arbitrary spectral index and declination can be calculated: while for hard spectra it is advantageous to suppress low energy events dominated by atmospheric neutrinos in order to increase the discovery potential, with larger declinations the increasing neutrino absorption may play an important role above 10 - 100 TeV on defining the optimal reconstructed muon energy threshold.

Based on this method, not only the best sensitivity can be derived but the possibility of detailed study of the source spectrum is also opened, provided sufficient statistics avoiding the unfolding of the measured muon spectrum. 
We apply the recipee from ~\citet{neronov09b} for the
calculation of the event rate in terms of reconstructed muon instead of neutrino energies. This approach presents several advantages including the possibility of exploiting some of the physical observables (such as reconstructed energy) for the optimization of IceCube performance for the detection of sources with different spectral characteristics.

\subsection{IceCube ``Spectral Discovery Curves''}

A major challenge for the detection of astronomical neutrino sources with IceCube is the separation of the source signal from the atmospheric neutrino background. If the spectrum of the source is harder than that of the atmospheric neutrino background, much of this background could be suppressed by selecting only higher-energy events, {\it i.e.} adjusting the reconstructed low muon energy threshold in the data analysis event selection. The optimal reconstructed low muon energy threshold, which maximizes the signal-to-noise ratio, depends, in general, on the shape (e.g. on the spectral index, for a power law neutrino energy distribution) of the spectrum of the source neutrinos. The dependence of the optimal analysis parameters on the spectral properties of the source leads to a dependence of the detector sensitivity on the assumptions about the spectral properties of the source. As a result, the knowledge of the IceCube discovery limit (= normalization of the neutrino spectrum at a given reference energy) for a "reference" source spectral index
does not allow to estimate the sensitivity limit for a different source spectrum. 

\begin{figure}
\centerline{\includegraphics[width=\linewidth]{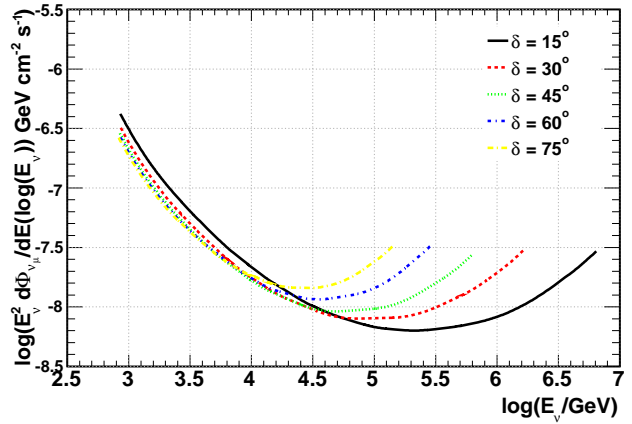}}
\caption{IceCube point source reach after three years of exposure. The tangent to one of these curves intersecting the $y$-axis at 1 GeV defines the muon neutrino flux normalization for the corresponding declination.}
\label{fig:Envelope-IC3yr} 
\end{figure}

In Appendix  \ref{sec:append} we demonstrate a convenient way of presenting the IceCube discovery limits for arbitrary assumptions about the neutrino spectrum slope of the source. The ``spectral discovery curves'' shown Fig. \ref{fig:Envelope-IC3yr} for various source declination bear the following meaning: 
A neutrino flux with spectral index $\Gamma_\nu$ (corresponding to a straight line in the $\log(E)$ vs. $E^2 \log(\mbox{ Flux})$ representation) tangent to one of the spectral discovery curve is at the limit of discovery.
%
%
This representation of the IceCube discovery limit(s) is useful, because it enables
\begin{itemize} 
\item The estimate of the IceCube sensitivity for arbitrary assumptions about the slope of the neutrino spectrum (i.e. the normalization of the minimal detectable spectrum at a given reference energy)
\item The estimate of the neutrino energy range contributing most significantly to the source signal (the energy at which the minimal detectable source spectrum touches the discovery curve).
\end{itemize}


The dependence of  the discovery curves, Fig.  \ref{fig:Envelope-IC3yr}, on the source location in the sky is reduced to the source declination  $\delta$, given the privileged location of IceCube at the South Pole.
We notice a weak dependence of the discovery limit on the source declination in the case of soft neutrino sources ($\Gamma_\nu \gg 2$): Most events contributing to the source signal have relatively low energies, at which Earth is transparent for neutrinos. 
For sources with hard spectra, $\Gamma_\nu \ll 2$, the IceCube discovery potential is strongly affected by the increase of the source declination.
The rise of the Earth absorption probability with energy obscures the source signal, contributed mostly by high energy neutrinos.
The effect is most dramatic for the hardest neutrino spectra, $\Gamma_\nu\simeq 1$. In this case, the minimal detectable flux for the sources at the declination $\delta=75^\circ$ is about 1.5 orders of magnitude higher than from sources at $\delta=15^\circ$.

\section{Discussion: IceCube potential for the study of the origin of high energy blazar activity}
\label{sec:hadronic}

The IceCube sensitivity curves, shown in Fig. \ref{fig:Envelope-IC3yr} can be directly applied to estimate the detectability of neutrino emission from {\it Fermi} blazars both in the model  with $\pi^0$ decay origin of the \gr\ spectrum and with the \gr\ emission from the secondary $e^+e^-$  pairs. 
%
We demonstrate this on the example of the model calculations for the TeV \gr\ loud blazar PG 1553+11.
The red thick curves in Figs. \ref{fig:e-2} and \ref{fig:e-1}, marked "IceCube", show the IceCube discovery curves for the source declinations $\delta=15^\circ$ (close to the actual declination of PG 1553+11) and $\delta=75^\circ$. 
The expected neutrino spectrum (thick green long-dashed line) clearly intersects the IceCube discovery curves in both figures. 
Therefore, if the source activity is determined solely by acceleration and subsequent interactions of protons,
the source should be detectable independent of the assumptions made about the origin of the GeV band \gr\  flux:
The detection or non-detection of several blazars with IceCube will confirm or rule out the so-called "hadronic" models of blazar activity.

Note that, although the expected neutrino signal from PG 1553+11 in Figs. \ref{fig:e-2} and \ref{fig:e-1} were calculated within the framework of a particular model of high energy proton interaction in the source (interaction with low energy protons from a radiatively inefficient accretion flow) developed in Ref. \citep{neronov09}, the numerical estimates of the overall flux and spectral characteristics of the neutrino signal from the sources are, in a sense, model independent. 

Indeed, on the one hand, in all hadronic models of blazar activity, which assume the high energy \gr\ emission  dominated by the emission from the neutral pion decay, the normalization and the spectral characteristics of the neutrino flux are constrained by the measurements of the \gr\ flux in the GeV band in a similar way. 

On the other hand, in all hadronic models which assume that the GeV band \gr\ emission is dominated by synchrotron and/or inverse Compton and/or Bremsstrahlung emission from the $e^+e^-$ pairs produced in the charged pion decay, the measurement of the GeV band \gr\ flux imposes a lower bound on the neutrino luminosity of the source (the neutrino luminosity should be larger or equal to the measured \gr\ luminosity). The shape of the neutrino spectrum at energies much below the threshold or low energy proton spectrum cutoff does not depend on the details of the proton spectrum itself: 
It is determined by the energy dependence of the differential cross section of proton interaction and charged pion decay spectrum. 
This means that in any model with \gr\ emission from secondary $e^+e^-$ pairs the expected neutrino flux is at least as high as the one shown in Fig. \ref{fig:e-1} and therefore detectable with IceCube, provided a low energy cutoff of the proton spectrum or the threshold energy of proton-photon interaction not higher than $\sim 10^{18}$~eV (see Fig. \ref{fig:e-1}).


To summarize, we have explored the IceCube potential introducing a generic method to derive power-law type spectra discovery thresholds for an arbitrary neutrino spectral index ($1\le\Gamma_\nu\le 4$) and for an arbitrary declination. In the light of {\it Fermi}, the IceCube potential in the search for blazar neutrino point sources is promising even for quite different types of hadronic models of activity: After three years of exposure,  several sources should be revealed independently of the details of the hadronic models. Detection or non detection of the TeV sources PG 1553+11 and Mrk421 with the data collected by the IceCube collaboration between 2007 and 2009 would already set strong constraints on the model with the soft spectrum of proton injection.


Moreover, we established that IceCube will provide a definite answer regarding the validity of the hadronic class of models of activity, ruling them out in the case of no discovery. 

In contrast, the first high energy neutrino source detections with IceCube would not only open a new observational window, IceCube would delineate the two classes of hadronic models of activity as well, from a concurrent study of the Fermi photon and IceCube neutrino indices: 
While detected blazars conforming to models predicting soft neutrino spectra will lie in the lower part of the diagram Fig. \ref{fig:diagram} (delimited by the IceCube curves), the ones conforming to models with hard neutrino spectra will be distributed over a large range of photon indices $\Gamma_\gamma$.
The detection of blazars with IceCube would therefore truly marks the birth of multimessenger high energy neutrino astronomy.


\section*{Acknowledgment}
The authors thanks L. Demir\"ors and the anonymous referee for valuable comments. M. Ribordy is supported by the Swiss National Research Foundation (grant PP002--114800).

\appendix
\section{Derivation of the IceCube spectral discovery curves}
\label{sec:append}

The neutrino point source strength normalization for discovery as a function of the neutrino source spectral index can be derived from the calculations of the number of atmospheric and extra-terrestrial neutrino-induced muon events.
These limits turn out to critically depend on the angular and energy resolution functions of the apparatus
due to the diffuse nature of the soft atmospheric neutrino background. 
The relative importance of the resolution functions depend strongly on the spectral shape of the point source neutrino flux. 
The sensitivity to hard spectra with cutoff energy above $\sim$ PeV will predominantly depend on the energy resolution while soft spectra will be probed best with instruments showing improved angular resolutions
(Existing instruments deployed either in ice or water do not combine optimal resolution functions, which largely depend on the medium properties. This holds to the Cherenkov photon ``thermalization'' power of the media, determined by the relationship between Cherenkov light absorption and scattering lengths:  In ice, $\lambda_\mathrm{abs} \gg \lambda_\mathrm{scat}$, in water $\lambda_\mathrm{abs} \simeq \lambda_\mathrm{scat}$. Therefore, while instruments using ice as the Cherenkov photon transport medium show improved calorimetric properties, instruments deployed in water have better pointing capabilities for opposite reasons)
.

Our simplified calculation for the required number of observed events pointing back to a source for a 5$\sigma$ C.L. discovery is based on Poisson probability, requiring $C_P(n_{\rm{obs}}-1,n_b) \ge \rm{erf}(5/\sqrt{2})$ where $n_s$ ($n_b$) is the number of signal (atmospheric) neutrino events, $n_{\rm{obs}}=n_s+n_b$ and $C_P$ is the Poisson cumulative distribution.

\begin{figure}
\centerline{\includegraphics[width=\linewidth]{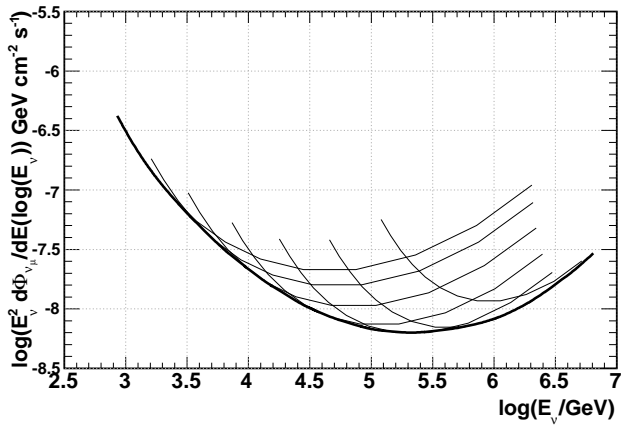}}
\caption{Power law limit representation of IceCube discovery potential. The thin curves (from left to right) corresponds to increasing reconstructed muon energy thresholds, from 100 GeV to 100 TeV, in step of 0.5 in log(E), for a declination $\delta=15^\circ$. The thick curve is the envelope, representing the best possible reach following the selection method of reconstructed muon energy threshold.}
\label{fig:envelopes} 
\end{figure}

For neutrino fluxes with spectral indices between $1 \le \Gamma_\nu \le 4$,
\begin{equation}
\frac{{\rm{d}}\Phi_{\nu_\mu}(E_\nu)}{{\rm{d}}E_\nu} = \Phi_0 \times \big(\frac{E_\nu}{\mathrm{GeV}}\big)^{-\Gamma_\nu} 
\end{equation}
we calculate the ``5$\sigma$ discovery'' flux normalization for a series of reconstructed muon energy thresholds, $E_\mu^{min}$. Fig.~\ref{fig:envelopes} illustrates the procedure leading to the derivation of the optimal potential by extraction of the envelope of curves from $\{ \log{(E_\nu^2 \frac{\mathrm{d}\Phi_\nu}{\mathrm{d}E_\nu}(E_\nu,\delta,E^{\mathrm{min}}_{\mu}))} \}_{E^{\mathrm{min}}_{\mu}}$, calculated for the sources at different declinations $\delta$. The meaning of each curve is the following: A power law neutrino spectrum with normalization tangent to the thick curve is at the discovery limit.

Technically, the extraction of the envelope is performed in two steps instead of the single one as illustrated in Fig~\ref{fig:envelopes}: once the best flux normalization $\Phi_0^{\mathrm{min}}(\Gamma_\nu,\delta)$ determined (the lower envelope of $\{ \log{\Phi_0} (\Gamma_\nu, \delta,{E^{\mathrm{min}}_{\mu}}) \}_{E^{\mathrm{min}}_{\mu}}$),
the smooth thick curve from Fig~\ref{fig:envelopes} is obtained from the envelope extraction from the series of curves  $\{\log{\Phi_0^{\mathrm{min}}(\Gamma_\nu,\delta)}\}_{\Gamma_\nu}$. The envelope condition reads
\begin{equation}
\log{E_\nu} = \frac{\mathrm{d}\log{\Phi_0^{\mathrm{min}}}}{\mathrm{d}\Gamma_\nu}(\Gamma_\nu,\delta),
\end{equation}
thus
\begin{equation}
\log{(E_\nu^2 \frac{\mathrm{d}\Phi_\nu}{\mathrm{d}E_\nu}(E_\nu,\delta))} =  \log{\Phi_0^{\mathrm{min}}(\Gamma_\nu,\delta)} + (2-\Gamma_\nu)\log{E_\nu}.
\end{equation}

The results are shown Fig.~\ref{fig:Envelope-IC3yr} for various declinations and represent the IceCube discovery potential to power law flux spectra. 


\end{document}